\documentclass[aps,prd,twocolumn,superscriptaddress,showpacs]{revtex4}

\usepackage{graphicx}  
\usepackage{dcolumn}   
\usepackage{bm}        
\usepackage{amssymb}   

\usepackage{xspace}

\def\etal{{\sl et al.}}
\def\etjj{{e\tau 2}}
\def\mtjj{{\mu\tau 2}}
\def\ltjj{{\ell\tau 2}}
\def\mt0{{\mu\tau 0}}

\newcommand{\met}{\mbox{\ensuremath{\slash\kern-.7emE_{T}}}}
\newcommand{\mht}{\mbox{\ensuremath{\slash\kern-.7emH_{T}}}}
\newcommand{\mpt}{\mbox{\ensuremath{\slash\kern-.5emT_{T}}}}

\newcommand{\ttbar}{\mbox{\ensuremath{t\overline t}}}
\newcommand{\wj}{\mbox{\ensuremath{W+\mathrm{jets}}}}
\newcommand{\zj}{\mbox{\ensuremath{Z+\mathrm{jets}}}}

\begin{document}



\begin{flushleft}

{FERMILAB-PUB-12-076-E} \\

\end{flushleft}

\title{Search for the standard model Higgs boson in tau lepton final states}


%
\affiliation{LAFEX, Centro Brasileiro de Pesquisas F\'{i}sicas, Rio de Janeiro, Brazil}
\affiliation{Universidade do Estado do Rio de Janeiro, Rio de Janeiro, Brazil}
\affiliation{Universidade Federal do ABC, Santo Andr\'e, Brazil}
\affiliation{University of Science and Technology of China, Hefei, People's Republic of China}
\affiliation{Universidad de los Andes, Bogot\'a, Colombia}
\affiliation{Charles University, Faculty of Mathematics and Physics, Center for Particle Physics, Prague, Czech Republic}
\affiliation{Czech Technical University in Prague, Prague, Czech Republic}
\affiliation{Center for Particle Physics, Institute of Physics, Academy of Sciences of the Czech Republic, Prague, Czech Republic}
\affiliation{Universidad San Francisco de Quito, Quito, Ecuador}
\affiliation{LPC, Universit\'e Blaise Pascal, CNRS/IN2P3, Clermont, France}
\affiliation{LPSC, Universit\'e Joseph Fourier Grenoble 1, CNRS/IN2P3, Institut National Polytechnique de Grenoble, Grenoble, France}
\affiliation{CPPM, Aix-Marseille Universit\'e, CNRS/IN2P3, Marseille, France}
\affiliation{LAL, Universit\'e Paris-Sud, CNRS/IN2P3, Orsay, France}
\affiliation{LPNHE, Universit\'es Paris VI and VII, CNRS/IN2P3, Paris, France}
\affiliation{CEA, Irfu, SPP, Saclay, France}
\affiliation{IPHC, Universit\'e de Strasbourg, CNRS/IN2P3, Strasbourg, France}
\affiliation{IPNL, Universit\'e Lyon 1, CNRS/IN2P3, Villeurbanne, France and Universit\'e de Lyon, Lyon, France}
\affiliation{III. Physikalisches Institut A, RWTH Aachen University, Aachen, Germany}
\affiliation{Physikalisches Institut, Universit\"at Freiburg, Freiburg, Germany}
\affiliation{II. Physikalisches Institut, Georg-August-Universit\"at G\"ottingen, G\"ottingen, Germany}
\affiliation{Institut f\"ur Physik, Universit\"at Mainz, Mainz, Germany}
\affiliation{Ludwig-Maximilians-Universit\"at M\"unchen, M\"unchen, Germany}
\affiliation{Fachbereich Physik, Bergische Universit\"at Wuppertal, Wuppertal, Germany}
\affiliation{Panjab University, Chandigarh, India}
\affiliation{Delhi University, Delhi, India}
\affiliation{Tata Institute of Fundamental Research, Mumbai, India}
\affiliation{University College Dublin, Dublin, Ireland}
\affiliation{Korea Detector Laboratory, Korea University, Seoul, Korea}
\affiliation{CINVESTAV, Mexico City, Mexico}
\affiliation{Nikhef, Science Park, Amsterdam, the Netherlands}
\affiliation{Radboud University Nijmegen, Nijmegen, the Netherlands}
\affiliation{Joint Institute for Nuclear Research, Dubna, Russia}
\affiliation{Institute for Theoretical and Experimental Physics, Moscow, Russia}
\affiliation{Moscow State University, Moscow, Russia}
\affiliation{Institute for High Energy Physics, Protvino, Russia}
\affiliation{Petersburg Nuclear Physics Institute, St. Petersburg, Russia}
\affiliation{Instituci\'{o} Catalana de Recerca i Estudis Avan\c{c}ats (ICREA) and Institut de F\'{i}sica d'Altes Energies (IFAE), Barcelona, Spain}
\affiliation{Uppsala University, Uppsala, Sweden}
\affiliation{Lancaster University, Lancaster LA1 4YB, United Kingdom}
\affiliation{Imperial College London, London SW7 2AZ, United Kingdom}
\affiliation{The University of Manchester, Manchester M13 9PL, United Kingdom}
\affiliation{University of Arizona, Tucson, Arizona 85721, USA}
\affiliation{University of California Riverside, Riverside, California 92521, USA}
\affiliation{Florida State University, Tallahassee, Florida 32306, USA}
\affiliation{Fermi National Accelerator Laboratory, Batavia, Illinois 60510, USA}
\affiliation{University of Illinois at Chicago, Chicago, Illinois 60607, USA}
\affiliation{Northern Illinois University, DeKalb, Illinois 60115, USA}
\affiliation{Northwestern University, Evanston, Illinois 60208, USA}
\affiliation{Indiana University, Bloomington, Indiana 47405, USA}
\affiliation{Purdue University Calumet, Hammond, Indiana 46323, USA}
\affiliation{University of Notre Dame, Notre Dame, Indiana 46556, USA}
\affiliation{Iowa State University, Ames, Iowa 50011, USA}
\affiliation{University of Kansas, Lawrence, Kansas 66045, USA}
\affiliation{Kansas State University, Manhattan, Kansas 66506, USA}
\affiliation{Louisiana Tech University, Ruston, Louisiana 71272, USA}
\affiliation{Boston University, Boston, Massachusetts 02215, USA}
\affiliation{Northeastern University, Boston, Massachusetts 02115, USA}
\affiliation{University of Michigan, Ann Arbor, Michigan 48109, USA}
\affiliation{Michigan State University, East Lansing, Michigan 48824, USA}
\affiliation{University of Mississippi, University, Mississippi 38677, USA}
\affiliation{University of Nebraska, Lincoln, Nebraska 68588, USA}
\affiliation{Rutgers University, Piscataway, New Jersey 08855, USA}
\affiliation{Princeton University, Princeton, New Jersey 08544, USA}
\affiliation{State University of New York, Buffalo, New York 14260, USA}
\affiliation{Columbia University, New York, New York 10027, USA}
\affiliation{University of Rochester, Rochester, New York 14627, USA}
\affiliation{State University of New York, Stony Brook, New York 11794, USA}
\affiliation{Brookhaven National Laboratory, Upton, New York 11973, USA}
\affiliation{Langston University, Langston, Oklahoma 73050, USA}
\affiliation{University of Oklahoma, Norman, Oklahoma 73019, USA}
\affiliation{Oklahoma State University, Stillwater, Oklahoma 74078, USA}
\affiliation{Brown University, Providence, Rhode Island 02912, USA}
\affiliation{University of Texas, Arlington, Texas 76019, USA}
\affiliation{Southern Methodist University, Dallas, Texas 75275, USA}
\affiliation{Rice University, Houston, Texas 77005, USA}
\affiliation{University of Virginia, Charlottesville, Virginia 22901, USA}
\affiliation{University of Washington, Seattle, Washington 98195, USA}
\author{V.M.~Abazov} \affiliation{Joint Institute for Nuclear Research, Dubna, Russia}
\author{B.~Abbott} \affiliation{University of Oklahoma, Norman, Oklahoma 73019, USA}
\author{B.S.~Acharya} \affiliation{Tata Institute of Fundamental Research, Mumbai, India}
\author{M.~Adams} \affiliation{University of Illinois at Chicago, Chicago, Illinois 60607, USA}
\author{T.~Adams} \affiliation{Florida State University, Tallahassee, Florida 32306, USA}
\author{G.D.~Alexeev} \affiliation{Joint Institute for Nuclear Research, Dubna, Russia}
\author{G.~Alkhazov} \affiliation{Petersburg Nuclear Physics Institute, St. Petersburg, Russia}
\author{A.~Alton$^{a}$} \affiliation{University of Michigan, Ann Arbor, Michigan 48109, USA}
\author{G.~Alverson} \affiliation{Northeastern University, Boston, Massachusetts 02115, USA}
\author{M.~Aoki} \affiliation{Fermi National Accelerator Laboratory, Batavia, Illinois 60510, USA}
% ...
%
%
\collaboration{The D0 Collaboration\footnote{with visitors from
$^{a}$Augustana College, Sioux Falls, SD, USA,
$^{b}$The University of Liverpool, Liverpool, UK,
$^{c}$UPIITA-IPN, Mexico City, Mexico,
$^{d}$DESY, Hamburg, Germany,
,
$^{e}$SLAC, Menlo Park, CA, USA,
$^{f}$University College London, London, UK,
$^{g}$Centro de Investigacion en Computacion - IPN, Mexico City, Mexico,
$^{h}$ECFM, Universidad Autonoma de Sinaloa, Culiac\'an, Mexico
and
$^{i}$Universidade Estadual Paulista, S\~ao Paulo, Brazil.
$^{\ddag}$Deceased.
}} \noaffiliation
\vskip 0.25cm

\date{March 20, 2012}

\begin{abstract}
We present a search for the standard model Higgs boson in final states
with an electron or muon and a hadronically decaying tau lepton in association with 
zero, one, or two or more jets using data corresponding to
an integrated luminosity of up to 7.3 fb$^{-1}$ 
collected with the D0 detector
at the Fermilab Tevatron collider.  The analysis is sensitive to
Higgs boson production via gluon gluon fusion, associated vector boson 
production, and vector boson fusion, and to Higgs boson decays to $\tau\tau$, $WW$, 
$ZZ$ and $b\overline b$ pairs. 
Observed (expected) limits are set on the ratio of 95\% C.L. upper limits on the 
cross section times branching ratio, relative to those predicted by the Standard Model, of
22 (14) at a Higgs boson mass of 115 GeV and 6.8 (7.7) at 165 GeV.
\end{abstract}

\pacs{13.85.Rm, 14.80Bn}

\maketitle


\section{Introduction}

The standard model (SM) of particle physics postulates a complex Higgs doublet field as the source of
electroweak symmetry breaking, giving rise to non-zero masses of the vector bosons and 
fundamental fermions.  The mass of the SM spin-zero Higgs boson, $H$, that survives after the symmetry
breaking is not predicted, but is constrained
by direct searches at the LEP~\cite{lep-higgs}, Tevatron~\cite{tevatron-higgs} and
LHC~\cite{lhc-higgs} colliders, to be in the range 115 -- 127 GeV at the 95\% C.L. 
Precision measurements of $W$ and $Z$ boson 
and top quark properties~\cite{gfitter} indicate a SM Higgs boson mass, 
$m_H = 96^{+31}_{-24}$~\cite{newwmass}.
Over the Higgs boson mass range $115 \leq m_H \leq 150$
the branching fractions 
vary considerably, with $H\rightarrow b\overline b$ ($H\rightarrow \tau^+ \tau^-$)  
being the dominant (subdominant) decays for $m_H \leq 135$ GeV
and $H\rightarrow W^+W^-$ ($H\rightarrow ZZ$) becoming important for $m_H > 135$ GeV.
Previous analyses by the D0 and CDF Collaborations have mainly 
focused on the decay modes $H\rightarrow b\overline b$ in the low mass region and 
$H\rightarrow WW$ with both $W$ bosons decaying to an electron or muon in the high mass
region.  

A previous D0 publication~\cite{ttjj_old} reported a Higgs boson search in 
the tau lepton pair plus two jets final state,
with one tau decaying to a muon and the other to hadrons, using 1.0 fb$^{-1}$ of data.  
The CDF collaboration has recently reported a similar search in the tau lepton pair plus at least
one jet~\cite{cdf_tautau}.
In this Letter, we report the results of three searches involving the production
of tau leptons 
that extend the previous results by adding more data, increasing the trigger efficiency, 
adding new search channels, and considering
additional signal contributions.  The final states used are:  
({\it i})  $\mu \tau$ plus zero or one jet (denoted $\mt0$), 
({\it ii})  $\mu \tau$ plus two or more jets ($\mtjj$), and 
({\it iii}) $e \tau$ plus two or more jets ($\etjj$).
The $\mt0$, $\mtjj$, and $\etjj$ analyses use 
data collected with the D0 detector~\cite{dzerodet} corresponding
to integrated luminosities of 7.3, 6.2 and 4.3 fb$^{-1}$ respectively.
The $e\tau0$ final state is not considered here as it suffers 
from large background and brings little increase in sensitivity.

\begin{figure}[htbp]
\includegraphics[scale=0.40]{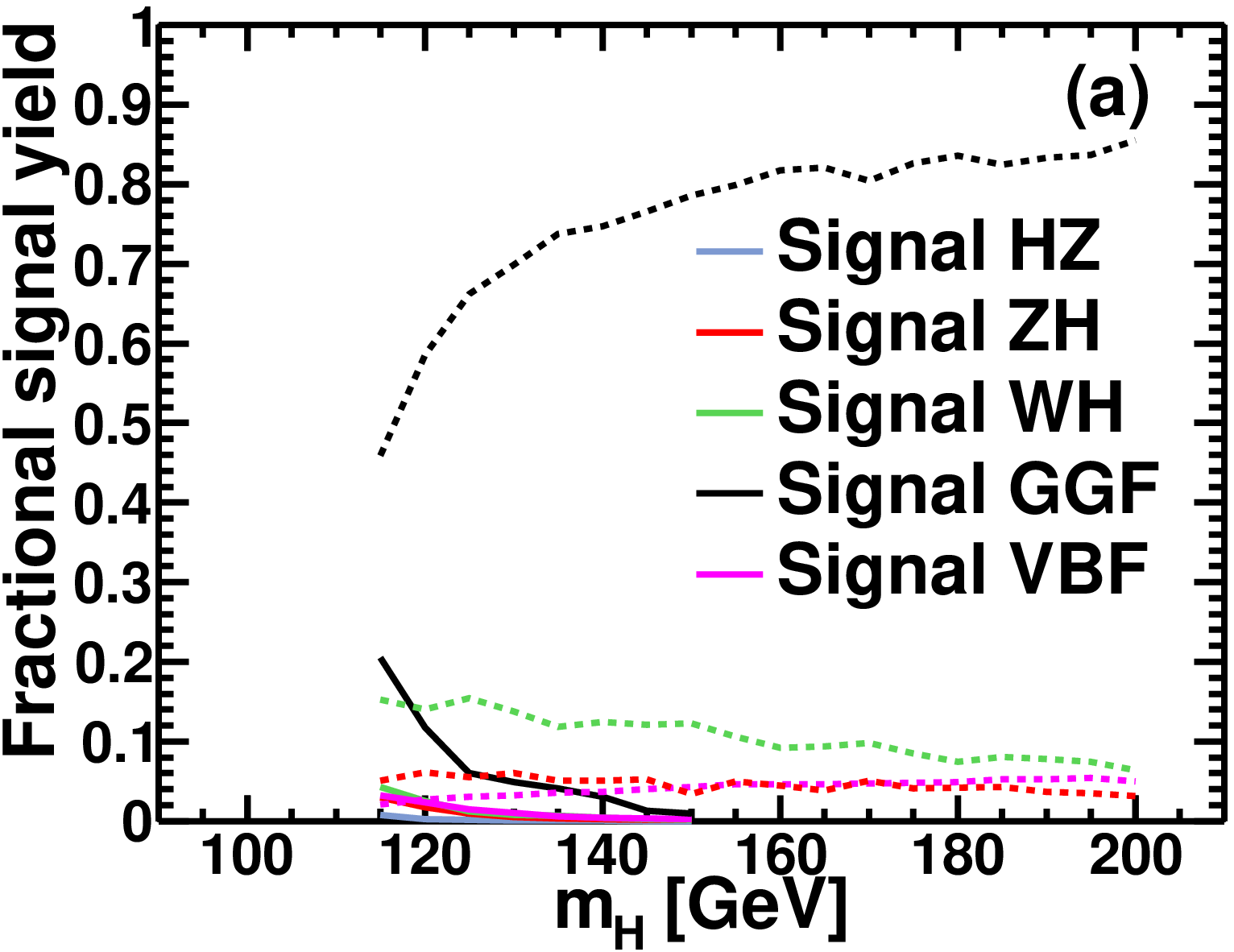}
\includegraphics[scale=0.40]{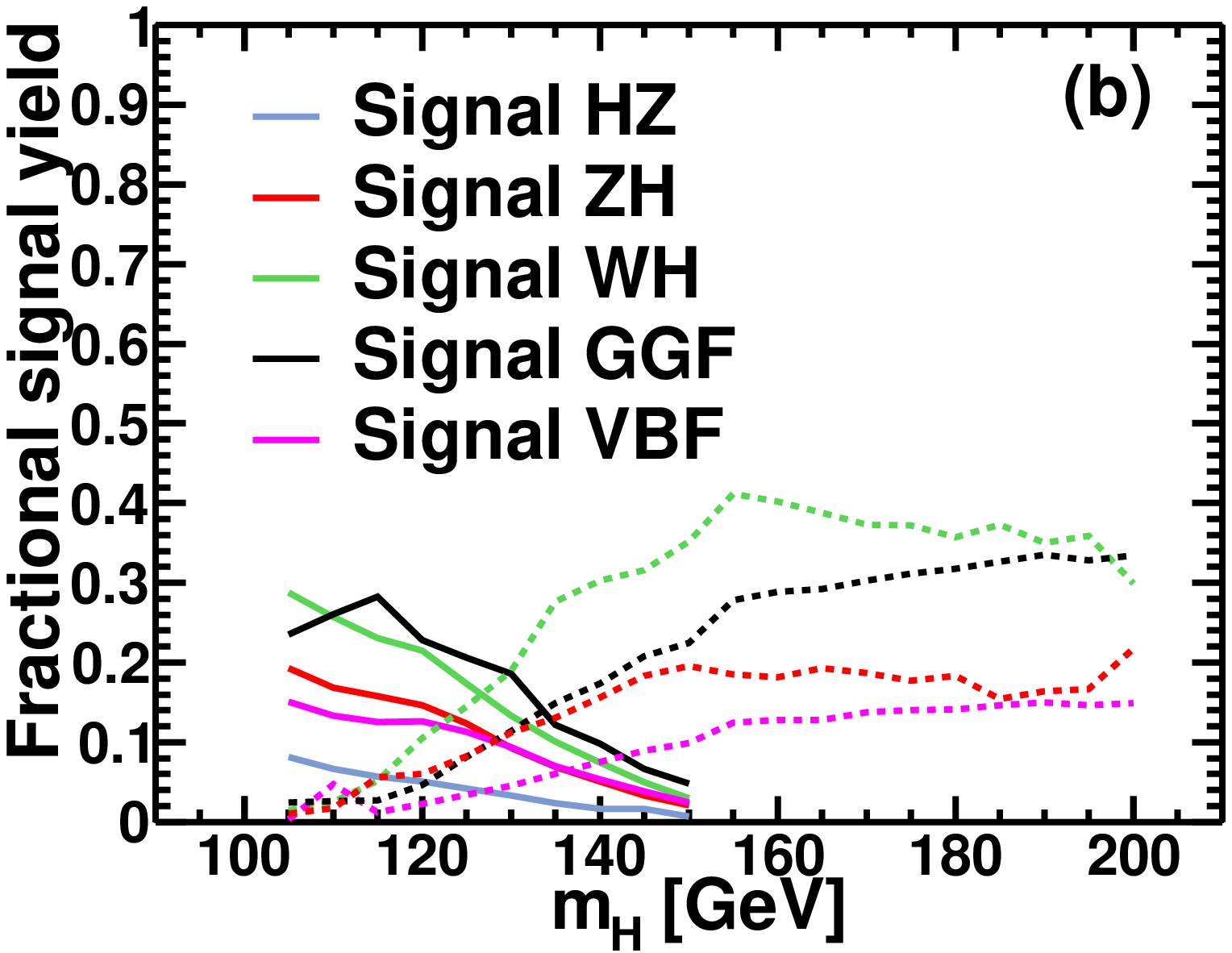}
\caption{\label{fig-signalyields}
(color online)
Fractional yields for $H$ signals from MC simulations
as a function of $M_H$ for (a) the $\mt0$ and (b) $\ltjj$ analyses. 
The yields for each signal process are plotted as solid lines for 
$H\rightarrow \tau\tau$ decays and as dashed lines
for the $H\rightarrow VV$ decays. 
}
\end{figure}

The Higgs boson production processes considered are 
({\it i}) gluon gluon fusion (GGF), $gg \rightarrow H$ (+ jets);
({\it ii}) vector boson fusion (VBF), $q \overline q \rightarrow q\overline q H$;
({\it iii}) associated vector boson and Higgs boson production (VH),
$q \overline q \rightarrow VH$, where $V$ is a $W$ or $Z$ boson, and $V\rightarrow q\overline q$
(or $Z \rightarrow \nu\nu$ in the case of $\mt0$); and
({\it iv}) associated Higgs boson and $Z$ boson production (HZ),
$q \overline q \rightarrow HZ$, with $H\rightarrow b\overline b$ and $Z\rightarrow \tau\tau$.
The GGF, VBF, and VH processes are further subdivided according to the Higgs boson
decay, $H\rightarrow \tau\tau$, $H\rightarrow WW$, or (for the $\mt0$ analysis) 
$H\rightarrow ZZ$, and these subchannels are denoted as GGF$_{\tau\tau}$, GGF$_{WW}$ or
GGF$_{ZZ}$, etc.
The fractional decompositions of signal contributions 
expected from Monte Carlo (MC) simulations
are shown in Fig.~\ref{fig-signalyields} for the Higgs boson production
cross section and branching ratios, and the event selection requirements, discussed below.  


Tau leptons can occur either through direct decays of the Higgs boson (at low mass) or indirectly 
from $H\rightarrow VV $ with $V$ decays to $\tau$s (at high mass).
The leptons may arise from $\tau$ decay or (at high mass) directly from $V$ decay.
The $\ell\tau$ channel 
is more uniformly sensitive to Higgs boson
production over the full allowed mass range than are the dedicated
$H\rightarrow b\overline b$ or $H\rightarrow WW\rightarrow \ell\overline \ell \nu\nu$ analyses, 
thus improving the sensitivity of a combination of searches,
particularly in the intermediate mass region around 135 GeV.
In the following, ``$\tau$'' represents a hadronically decaying tau and
``lepton ($\ell$)'' denotes $e$ or $\mu$.

\section{Trigger }

The $\mt0$ and $\mtjj$ data were collected from the full suite of D0 triggers.  
The main contributors were the inclusive high transverse momentum muon, $\mu$ + jets, 
and $\mu + \tau$ triggers.  The trigger efficiency is determined in a two-step 
procedure starting from the measurement of the efficiency for inclusive muon triggers.  
This is measured using $Z\rightarrow \mu\mu$ candidates and parameterized as a function 
of muon transverse momentum ($p_T$), pseudorapidity ($\eta$), azimuthal angle ($\phi$), 
and instantaneous luminosity.  We then determine the ratio of the yields of the full 
trigger suite relative to those for the inclusive muon triggers.  For the $\mt0$ 
analysis the ratio is parametrized as a function of $p_T^\tau$ while for the $\mtjj$ 
analysis it is a constant.  The efficiency for $Z\rightarrow \tau\tau$ events for 
the full suite of triggers varies between 80\% and 95\%, and is about 40\% larger than 
for the inclusive muon trigger.

For the $\etjj$ analysis,  
a set of calorimeter-based inclusive electromagnetic object triggers was used.  
The efficiency of these triggers, obtained from an analysis of $Z\rightarrow ee$
events selected with just one identified electron,
is found to be about 85\%.

\section{Background and signal samples}

The major backgrounds for the Higgs boson search are $\zj$, $\wj$, $\ttbar$, and 
multijet production (MJ) with misidentification of leptons or taus.  Smaller
backgrounds arise from boson ($W, Z$ or $\gamma$) pair production 
and single top quark production.
All but the MJ background are simulated using 
MC event generator programs and normalized
to the highest available next-to-leading order (NLO) 
or next-to-NLO (NNLO) theoretical calculations.  
The MC simulations use
the CTEQ6L1 parton distribution functions (PDF)~\cite{cteq}.  

The $\zj$ and $\wj$ event samples are generated by {\footnotesize ALPGEN}~\cite{alpgen}, interfaced
to {\footnotesize PYTHIA}~\cite{pythia} which provides initial and final state radiation and 
hadronization of the produced partons. The $p_T^Z$ distribution is reweighted to agree
with the D0 measurement~\cite{d0zpt}.  The $p_T^W$ is also reweighted 
for the $\ltjj$ analyses using the reweighting factors 
derived for the $p_T^Z$ distribution, multiplied by the ratio of the 
$p_T^W$ to the $p_T^Z$  distributions as predicted in 
NNLO QCD~\cite{nnlowpt}. For the $\ltjj$ analyses, 
the absolute normalization for the $\zj$ and $\wj$ cross sections
are taken from Ref.~\cite{nnlovjxs} using the MRST2004 NNLO PDFs~\cite{mrst2004}.
The same $\zj$ normalization is used for the $\mt0$ analysis but the $\wj$ normalization
is derived from data as discussed below.

We simulate $\ttbar$ and single top quark events using the {\footnotesize ALPGEN} 
and {\footnotesize COMPHEP}~\cite{comphep}
generators respectively, with {\footnotesize PYTHIA} used to simulate hadronization effects.  
The normalizations are based on 
the approximate NNLO calculations~\cite{topxs}.  The diboson events are generated
by {\footnotesize PYTHIA}.

Higgs boson production is simulated using {\footnotesize PYTHIA}, with normalizations
taken from Ref.~\cite{higgsxs}.  We use 
{\footnotesize HDECAY}~\cite{hdecay} and
{\footnotesize TAUOLA}~\cite{tauola} to obtain the branching fractions of the Higgs
boson and tau lepton respectively. 

All MC signal and background events are input to a 
{\footnotesize GEANT3}-based ~\cite{geant} simulation 
of the detector response and processed with the same reconstruction
programs as used for data.  Data events collected from random beam crossings are superimposed
on the MC events to account for detector noise and pileup from additional $p\overline p$ collisions in the
same or previous bunch crossings.  Correction factors are applied to the simulated events to
account for the trigger efficiencies and for the differences between MC and data for 
the lepton, tau, and jet identifications, and for the energy scale and resolution of jets.

\section{Event selection criteria}

Muons selected for this analysis are required to have hits in the muon chambers before 
and after the toroidal magnets and to be matched to a track in the 
tracking system with $p_T>15$ GeV and $|\eta|<1.6$.  
Muon candidates are required to be isolated in both the calorimeter
and the tracking system using the calorimeter transverse energy, $E_T^{\rm iso}$, in the annular cone
$0.1<$ $\cal R$ $<0.4$ around the muon, where $\cal R$ $=\sqrt{(\Delta\eta)^2 + (\Delta\phi)^2}$, 
and the track transverse momentum sum,
$p_T^{\rm iso} = \Sigma p_T^{\rm track}$, within a cone $\cal R$ $<0.5$, excluding the $p_T$ of 
candidate muon.  For the $\mt0$ analysis, $E_T^{\rm iso}$ and $p_T^{\rm iso}$ must be less than
15\% of $p_T^\mu$.  For the $\mtjj$ analysis, $E_T^{\rm iso}$ and $p_T^{\rm iso}$ must
be less than 2.5 GeV.  Muon candidates due to cosmic rays 
are rejected if the scintillation counters surrounding
the detector indicate a time of arrival different by 
more than 10 ns from that expected for collision products.

Electrons are identified using a likelihood variable, $\cal L$$_e$, that uses as inputs 
the quality of the matching of the electromagnetic (EM) shower centroid to a track, 
the fraction of energy deposited in the EM section  of the calorimeter (EMF), 
a measure of the probability that the energy deposit pattern in the 
calorimeter conforms to that expected for an electron, 
$E_T^{\rm iso}$, and the
separation along the beam axis of the electron track and the primary vertex (PV)~\cite{pv}.  
The signal sample electrons are required to have $\cal L$$_e > 0.85$.
Electron candidate tracks are required to have $p_T>15$ GeV and $|\eta| <1.1$
or $1.5 <|\eta| < 2.5$, and to impinge upon a module of the central EM calorimeter
within the central 80\% of its azimuthal range.

The selection of hadronically decaying tau leptons 
is done separately for three types based on the number of 
tracks within a cone $\cal R$ $<0.3$ and the number of EM subclusters found in
the calorimeter using a nearest neighbor algorithm.  Type-1, patterned on the decay
$\tau \rightarrow \pi \nu_\tau$, requires one track and no EM subclusters.  
Type-2, based on $\tau\rightarrow \rho (\pi^\pm \pi^0) \nu_\tau$, requires one track and at
least one EM subcluster.
Type-3, motivated by the $\tau\rightarrow 
\pi^\pm \pi^\pm \pi^\mp (\pi^0) \nu_\tau$ decay, requires at least two
tracks with or without EM subclusters.  We reject type-3 candidates 
with exactly two tracks of opposite
signs since their charge sign is ambiguous. 
The $\tau$ transverse energy, $E_T^{\tau}$, is defined as the visible transverse 
momentum of the $\tau$ decay products as measured by the calorimeter with appropriate 
energy scale corrections.  The ratio of $E_T^\tau$ to the sum of the tracks associated with 
the tau, $p_T^{\rm trk}$, is used to verify that the MC and data tau energy scales are the same.
We require $E_T^\tau > (12.5, 12.5, 15)$ GeV, 
$p_T^{\rm trk} > (7, 5, 10)$ GeV, and
$(p_T^{\rm trk} / E_T^\tau) > (0.65, 0.5, 0.5)$
for $\tau$ types (1, 2, 3).  The leading (highest $p_T$) track for type-3 $\tau$s must exceed 7 GeV.
A neural network, NN$_\tau$~\cite{taunn}, based on energy deposition patterns and isolation
criteria in the calorimeter and tracking systems is constructed for each tau type to discriminate a 
$\tau$ from a misidentified jet.  
Lower bounds placed on NN$_\tau$ at 0.9, 0.9 and 0.95 for tau types 1, 2, and 3 select hadronically
decaying taus with good purity.
For type-2 $\tau$ leptons we discriminate taus from electrons using 
a second neural network, NN$_{\tau/e}$, constructed using
variables that characterize the 
longitudinal and transverse energy profiles in the calorimeter, the energy
and position correlations between $\tau$ tracks and calorimeter energy deposits,
and isolation of the calorimeter energy.  

Jets are selected using an iterative midpoint cone algorithm~\cite{jetalg} with a cone size
$\cal R$ = 0.5.  
We require at least two tracks associated with the jet that point to the PV.
Jet energies are corrected to the particle level for out-of-cone showering, underlying event 
energy deposits and pileup, and the estimated missing energy in jets with identified semileptonic
decays of a hadron.     
The energy scale, resolution, and 
jet identification efficiency for MC jets are corrected to give agreement with data.
For the quark-dominated MC samples 
($t\overline t$ and diboson), there is an additional correction of the jet energy that accounts for
the differences in the responses of quark jets and the dominantly gluon jets for which the 
jet energy scale correction was obtained.
The $\mtjj$ and $\etjj$ analyses require at least two jets with $|\eta_{\rm jet}|<3.4$ and
$p_T^{\rm jet} > 20$ (15) GeV for the leading (other) jet.
The $\mt0$ analysis imposes these jet $p_T$ requirements as a veto 
to ensure that the selected samples have no events in common.

The missing transverse energy, 
$\met$, is computed from the observed transverse energy deposits in the calorimeter 
and is adjusted for 
the appropriate energy scale corrections for all objects and for isolated muons observed in the event.

For the final event selection, all three analyses require exactly one isolated lepton and a hadronic
tau with opposite charges.  The separations between all pairs of
lepton, tau, and jet are required to be $\cal R$ $> 0.5$.  
For the $\mt0$ analysis, events are required to have only one $\tau$, and the 
smaller of the transverse masses, 
$m_T = \sqrt{2E_T^{\rm lepton} ~\met (1-\cos\Delta\phi)}$ 
(where ``lepton" = $\tau$ or $\mu$ and $\Delta\phi$ is the angle between
the lepton and~$\met$) 
must exceed 25 GeV to suppress the $Z+$ jets 
and MJ backgrounds, while retaining about 80\% of the signal.
For the $\etjj$ analysis, substantial backgrounds arise from $\zj$ production with 
$Z \rightarrow ee$ where an electron is misidentified as a type-2 $\tau$.  
To reduce these, we remove $\tau$ candidates in the region 
$1.1 < |\eta| < 1.5$ where the calorimetry has impaired electron identification.  Further 
$Z+$ jets rejection is obtained by requiring type-2 $\tau$ candidates to have NN$_{\tau/e} > 0.95$
to suppress electrons that resemble the track + EM cluster signature.
This cut retains more than 80\% of
type-2 $\tau$s while rejecting about 90\% of the electrons~\cite{nnel}.
We reject type-2 $\tau$ candidates which point near the edge in $\phi$
of an EM module in the central calorimeter where the EM response is impaired.
In addition, type-3 $\tau$ candidates with EMF $>0.95$ are excluded.  The MJ background in
the $\etjj$ analysis is suppressed by requiring $\cal S$ $>1$, where $\cal S$ is 
a measure of the significance for ~$\met$ to differ from zero~\cite{metsig}.

\section{Backgrounds derived from data}

The MJ background arising from misidentification of leptons or taus
by the detector reconstruction algorithms is difficult to simulate, so for 
each analysis, the MJ background is taken from data.  The general method for
all analyses is similar: we define a sample of MJ-enriched events, $\cal M$, from which
residual backgrounds simulated by MC are subtracted, to provide the shapes of the MJ
kinematic distributions.  The number of MJ events in the signal sample is
obtained by multiplying the MJ yield in a signal-like sample $\cal N$
by a scale factor $\rho_i$, obtained from
the $\cal M$ sample for each of the tau types, $i$.  
The $\rho_i$ factors provide the estimate
for the differences in the MJ background normalization between the $\cal N$
and signal samples, based on the $\cal M$ sample, and are in all cases 
within 10\% of unity.

For the $\mt0$ channel, the sample $\cal M$ is obtained by requiring 
$m_T(\mu,\met) < 30$ GeV and NN$_\tau < 0.2$, and the $\rho_i$ are the ratios
of isolated to non-isolated lepton events in $\cal M$, and are
parameterized as a function of $p_T^\tau, N_{\rm jets}, ~\met$, and $p_T^\mu$.  These 
factors scale the MJ fraction of the sample $\cal N$, 
selected as for the signal sample except that the muon is
required to be non-isolated, 
to obtain the MJ normalization in the isolated lepton signal sample.
An alternate MJ-enriched sample is defined by NN$_\tau < 0.2$ and $m_T(\mu,\met) < 30$ GeV,
in which the $\tau$ and $\mu$ have the same charge sign, for estimating
the MJ background uncertainty. 

For the $\mtjj$ analysis, the MJ sample $\cal M$ is obtained by reversing at least
one of the muon isolation requirements and requiring $0.3 <{\rm NN}_\tau < 0.8$. 
The MJ fraction of this sample is 94\% before the MC-simulated background subtraction.
The $\rho_i$ factors are the ratio of opposite charge sign (OS) and same charge sign (SS) 
$\mu - \tau$ pairs in $\cal M$ and 
are used to scale the MJ component of the sample $\cal N$ selected as for the signal
sample except that we require SS $\mu$ and $\tau$.
The $\rho_i$ show no significant dependence on the kinematic
variables.

For the $\etjj$ analysis, $\cal M$ is obtained by requiring the electron to satisfy an 
orthogonal loose electron selection, $0 <$ $\cal L$$_e <0.85$, 
and $0.3 <{\rm NN}_\tau < 0.9$.  
The MJ fraction of this sample is 96\% before the MC-simulated background subtraction.
The $\rho_i$ are
obtained from the OS and SS $\cal M$ sample and
are applied to the MJ component of the $\cal N$ sample as in the $\mtjj$ analysis. 
The $\rho_i$ show
no significant dependence on kinematic variables.
Alternate MJ-enriched samples, in which either the $\tau$ or lepton selections (but not both) 
are reversed, are defined for estimating the MJ background uncertainties in both
$\ell \tau 2+$ analyses.

For the $\mt0$ analysis, the dominant background is from $\wj$ with the muon from
$W$ decay and a jet misidentified as a tau.  Both the normalization of the $\wj$ 
sample and the misidentification probability are difficult to model adequately,
so the simulation is corrected using a data-driven method~\cite{madar}.  The jet produced in association
with a $W$ boson has a charge that is correlated differently with the $W$ boson charge
for quarks and gluons.  Furthermore, the probability for 
a jet misidentified as a tau to have
the same charge sign as its progenitor parton varies with NN$_\tau$.  We determine
a weight for $\wj$ MC events that depends on the charge correlation between the muon
and recoil parton and on the value of NN$_\tau$.

\section{Event Yields}

The numbers of data and expected background events are given in
Table~\ref{tab-yield} for the 
$\mt0$, $\mtjj$, and $\etjj$ analyses.


\begin{table}[htbp]
\caption{\label{tab-yield}
For each analysis channel, 
the number of  background events expected from MC simulated processes, MJ background, 
and observed data, for individual and sum of all tau types   
after preselection. 
``$V$+j'' denotes $W$ or $Z$ + jets and ``DB'' denotes diboson processes. 
}

\begin{tabular}{crrrrrrrr} \hline \hline
$\tau$ type & ~~$\ttbar$ & ~$W$+j & $Z_{\ell \ell}$+j & $Z_{\tau\tau}$+j & DB~ 
    & MJ~ & $\Sigma$Bkd  & Data  \\ \hline 

\multicolumn{9}{c}{$\mt0$ analysis}  \\ \hline
 type 1    & 4  & 234  & 22 & 11 & 29 & 39 & 338  & 340  \\ 
 type 2    & 19 & 852  & 94 & 56 & 108  & 116  & 1245 & 1294 \\ 
 type 3    & 4  & 678  & 57 & 19 & 25 & 67 & 850  & 839  \\ 
 All       & 27 & 1764 & 172  & 86 & 162  & 223  & 2433 & 2473 \\ \hline

\multicolumn{9}{c}{$\mtjj$ analysis} \\ \hline
 type 1    & 13  & 9   & 5  & 29  & 2  & 19 & 76  & 81   \\ 
 type 2    & 86  & 57  & 22 & 159 & 12 & 59 & 394 & 418  \\ 
 type 3    & 13  & 34  & 4  & 43  & 3  & 22 & 119 & 109  \\ 
 All       & 112 & 100 & 31 & 231 & 16 & 99 & 589 & 608  \\ \hline

\multicolumn{9}{c}{$\etjj$ analysis}  \\ \hline
 type 1    & 2  & 2  & 0  & 6  & 1 & 6  & 18  & 10  \\ 
 type 2    & 14 & 21 & 14 & 30 & 2 & 25 & 106 & 98  \\ 
 type 3    & 7  & 16 & 2  & 11 & 1 & 15 & 52  & 59  \\ 
 All       & 24 & 40 & 16 & 46 & 4 & 47 & 176 & 167 \\ \hline \hline

\end{tabular}
\end{table}

\section{Multivariate Analysis}

\begin{figure*}[t]
\begin{center}
\includegraphics[width=0.330\textwidth]{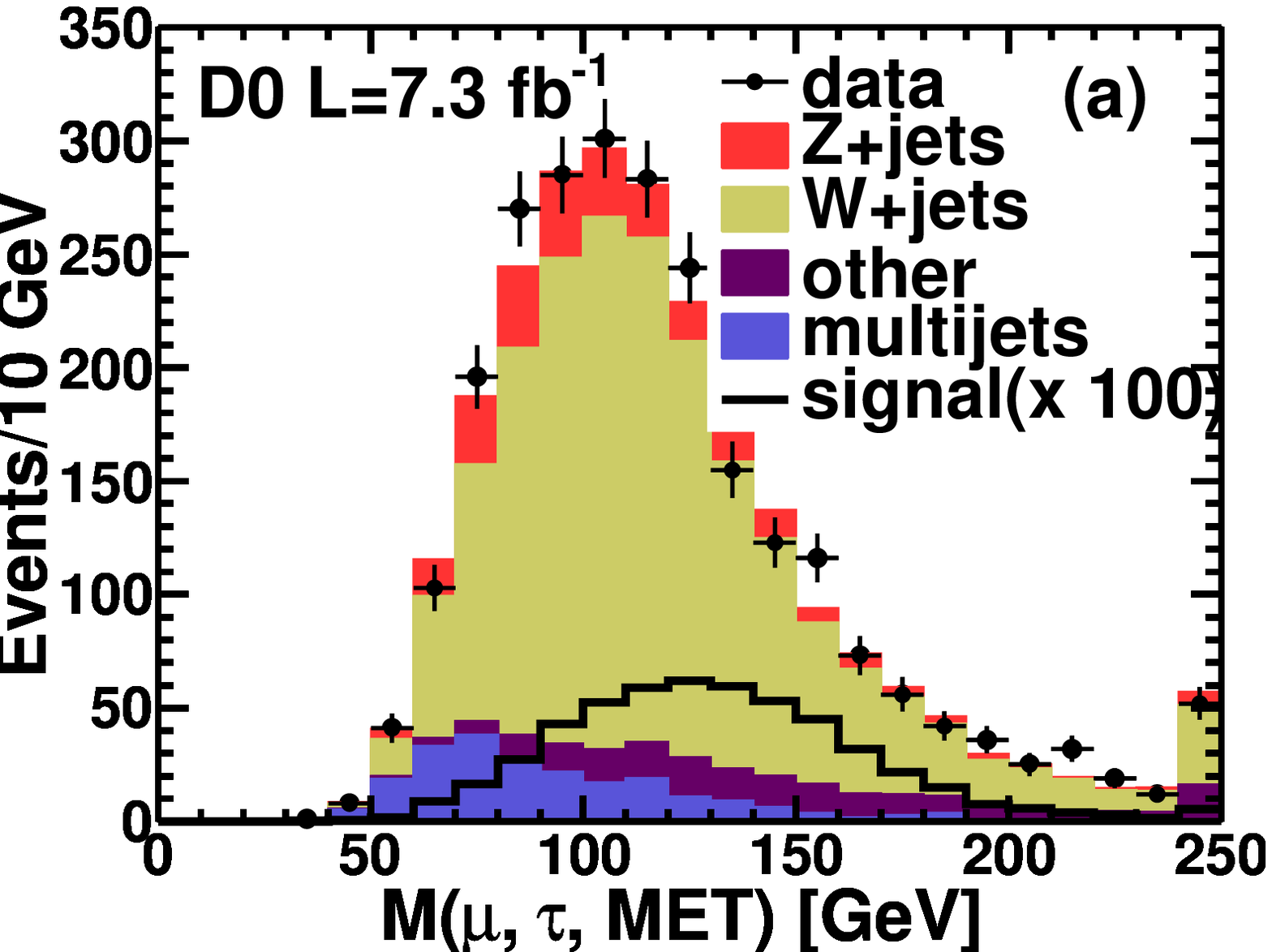}
\includegraphics[width=0.325\textwidth]{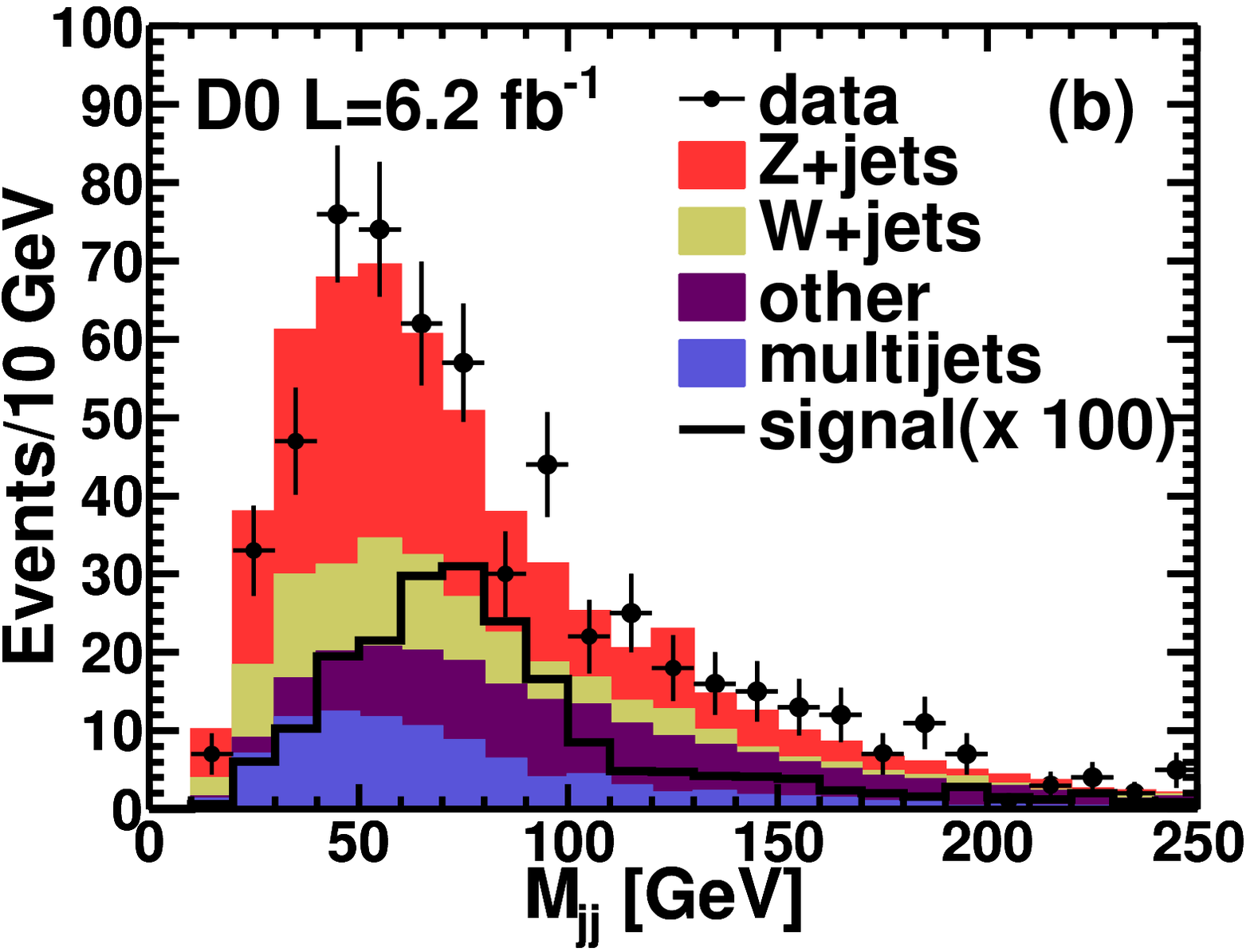}
\includegraphics[width=0.325\textwidth]{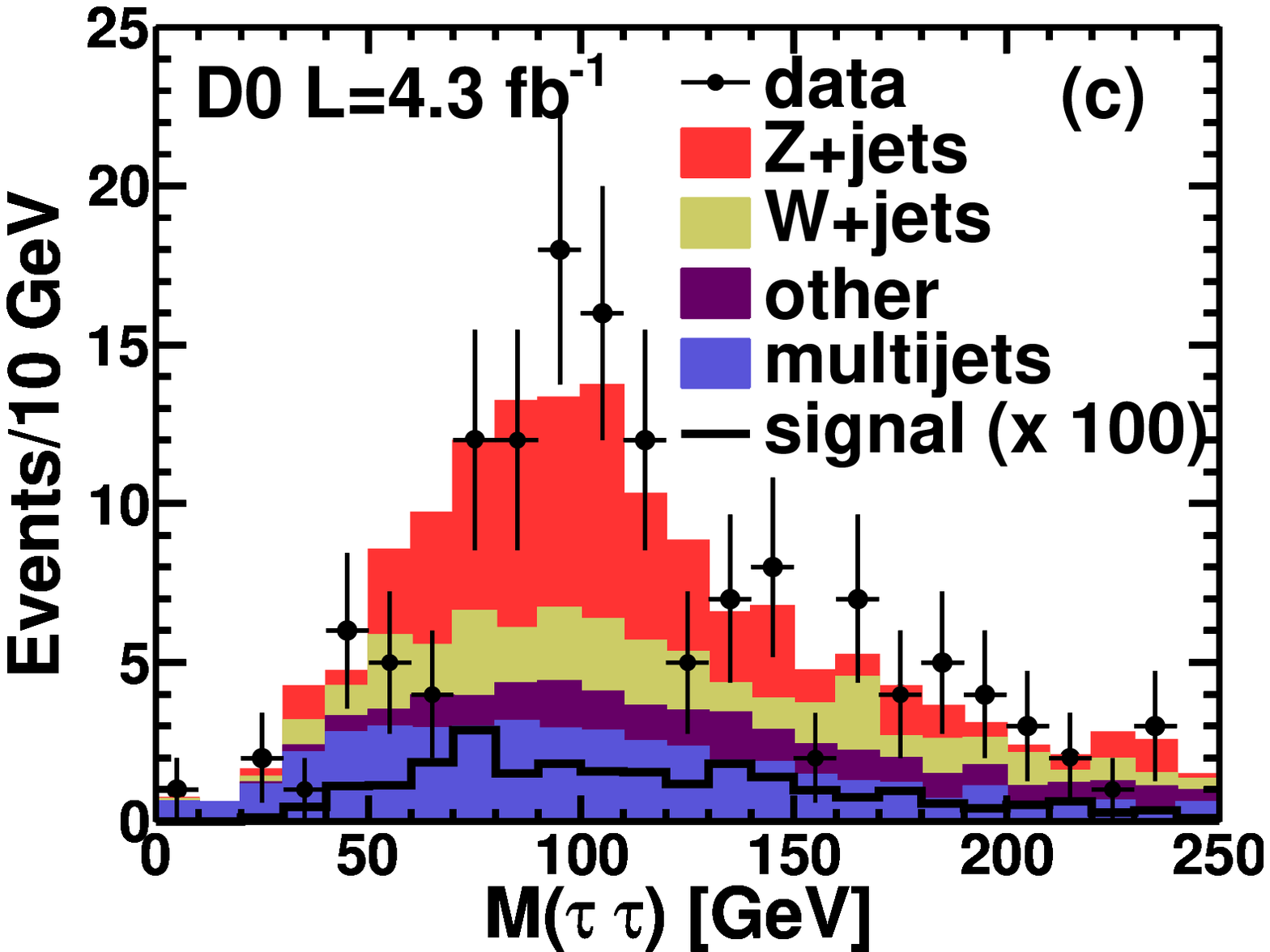} 
\caption{\label{fig_inputs}
(color online) Comparison of data, expected backgrounds, and total signal with the indicated
scaling factors for 
(a) invariant mass of the $\mu, \tau, ~\met$ system for the $\mt0$ analysis 
(signal shown for $m_H=165$ GeV);
(b) dijet invariant mass for the $\mtjj$ analysis
(signal shown for $m_H=150$ GeV); and
(c) invariant mass of the $\tau\tau$ system for the $\etjj$ analysis
(signal shown for $m_H=150$ GeV), where the ~$\met$ contribution
is apportioned to the $e$ and $\tau$ as discussed in the text.
The ``MET" in the labels refers to~$\met$.  
The $t\overline t$, single top and diboson backgrounds are 
shown together as ``other".
}
\end{center}
\end{figure*}

The expected number of events for Higgs boson signal processes is small in comparison
to the backgrounds shown in Table~\ref{tab-yield}.
For example, the expected signal yields at $m_H= 165$ GeV are 
5.2, 1.7 and 0.3 events for the $\mt0$, $\mtjj$ and $\etjj$ analyses respectively.  
The corresponding yields at $m_H=115$ GeV are 0.9, 1.6 and 0.4 events.
We thus employ multivariate techniques that utilize both the magnitudes of the variables 
and the correlations among them to separate the signal from the backgrounds.  
We choose well-modeled variables that have the
capability to distinguish between at least one signal and one background as shown
in Table~\ref{tab-vars}.  Figure~\ref{fig_inputs} shows distributions for representative
variables that offer significant discrimination of signal and background for 
each of the channels.
For calculating the $\tau\tau$ invariant mass shown in Fig.~\ref{fig_inputs}(c),
the $\met$ is apportioned to the neutrinos from the two postulated 
tau leptons by 
decomposing the $\met$ vector into components associated with the observed lepton and 
hadronic tau~\cite{collinear-met}.


\begin{table}[htbp]
\caption{\label{tab-vars}
List of variables used in the multivariate discriminants for the $\mt0$ and $\ltjj$
analyses.  The variable $\mht$ is the missing transverse energy computed from
the jets in the event.
}

\begin{tabular}{lcc} \hline \hline
Variable & $\mt0$ & $\ell\tau 2+$  \\ \hline
Lepton $p_T$                                                & x & x \\
Tau $p_T$                                                   & x & x \\
Leading jet $p_T$                                           & x & x \\
$\met$                                                      & x & x \\ 
$\mu$ charge $\times \eta_\mu$                              & x &   \\
$\eta_\tau$                                                 & x &   \\
$\ell\tau~\met$ invariant mass                              & x &   \\ 
$\tau\tau$ invariant mass~\cite{collinear-met}                                  &   & x \\   
Dijet invariant mass                                        &   & x \\  
$\mu~\tau~\met$ invariant mass                              & x &   \\   
$\ell \nu$ transverse mass, $m_T^\ell$                      &   & x \\    
$\tau \nu$ transverse mass, $m_T^\tau$                      &   & x \\  
Minimum of $m_T^\ell$ and $m_T^\tau$                        & x &   \\  
$\Sigma |\vec p_T|$ of all jets                             &   & x \\  
Scalar $p_T$ sum of $\ell, \tau, ~\met$, jets               &   & x \\  
Magnitude of vector $p_T$ sum of $\ell, \tau,~\met$, jets   &   & x \\  
Minimum $\sqrt s$ necessary for final objects               & x &   \\ 
Number of jets                                              & x &   \\ 
$\Delta$$\cal R$ between leading jets                       &   & x \\ 
$\Delta \eta$ between leading jets                          &   & x \\ 
Asymmetry between $\met$ and $\mht$                         &   & x \\ 
$\Delta \phi$ between $\ell$ and $\tau$                     & x &   \\ 
$\Delta \theta$ between $\ell$ and $\tau$                   & x &   \\ 
$\Delta \phi$ between $\ell$ and $\met$                     & x &   \\ 
$\Delta \phi$ between $\tau$ and $\met$                     & x &   \\ 
$\Delta \phi$ between $\met$ from calorimeter and tracks    &   & x \\ 
Cosine of angle between $\ell$ and beam direction           & x &   \\ 
Minimum $\delta\phi$ between $\met$ and a jet               &   & x \\ 
Missing $E_T$ significance, $\cal S$                        &   & x \\
NN$_\tau$                                                   & x &   \\ \hline \hline
 
\end{tabular}
\end{table}

\begin{figure*}[t]
\begin{center}
\includegraphics[width=0.329\textwidth]{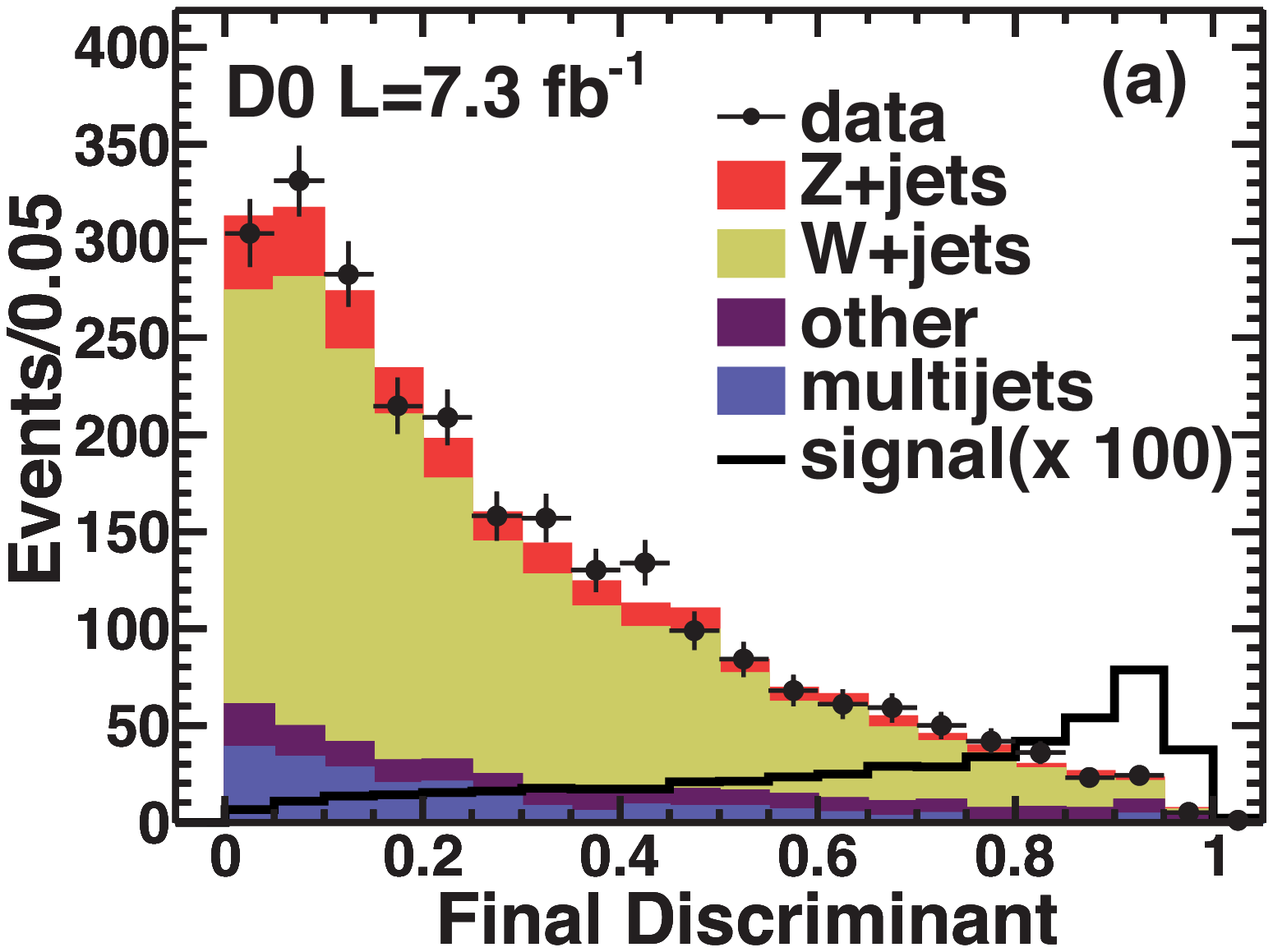}
\includegraphics[width=0.325\textwidth]{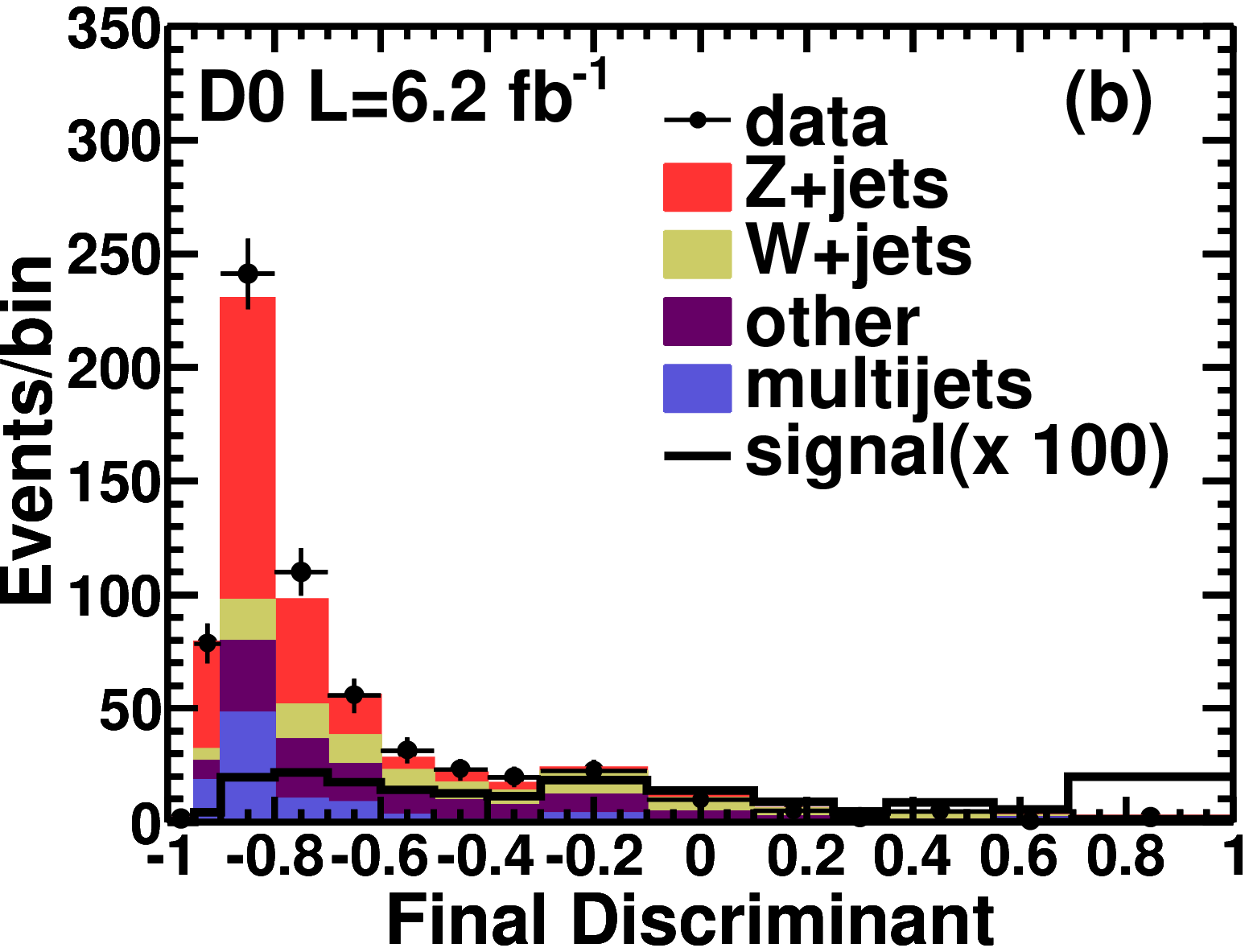}
\includegraphics[width=0.325\textwidth]{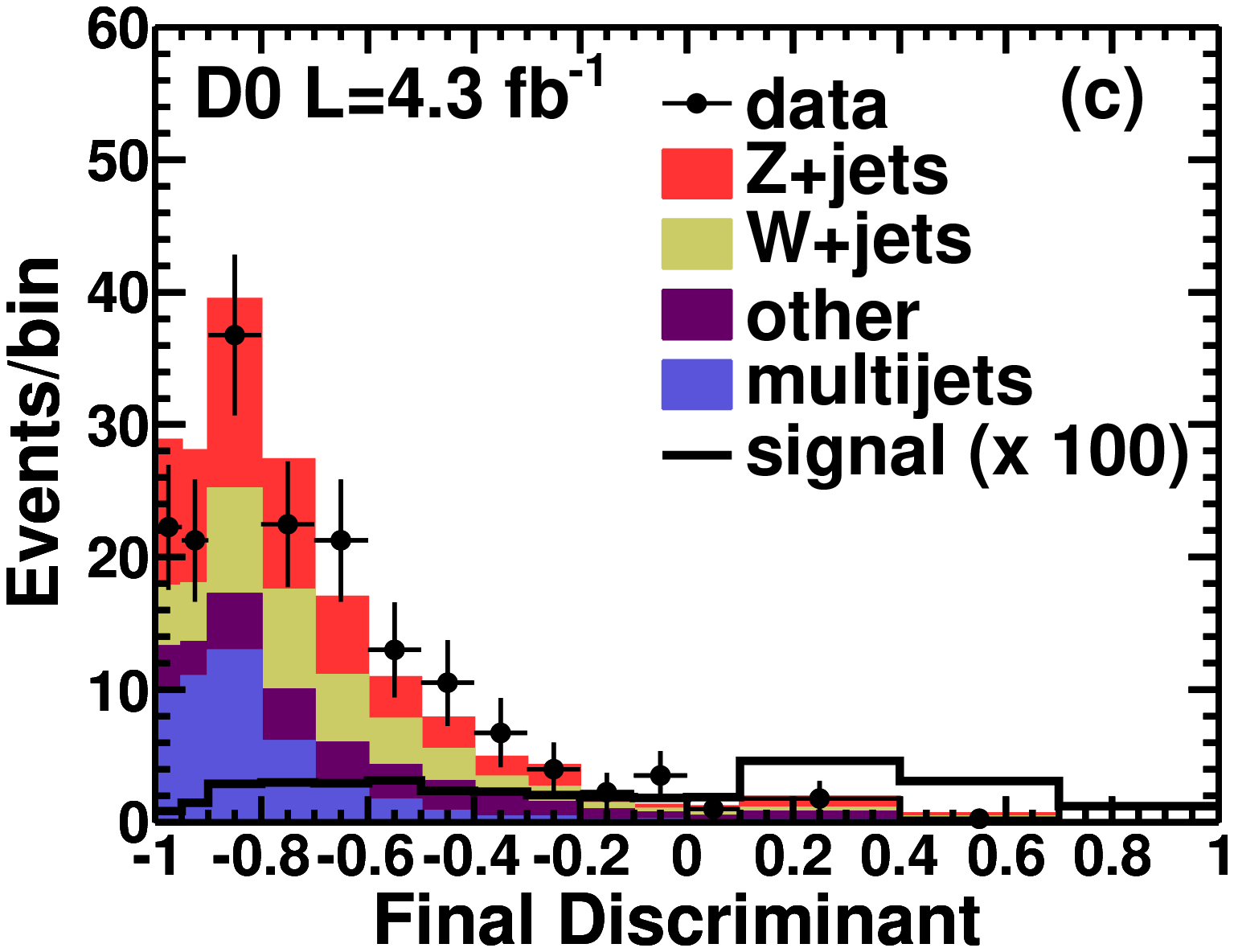} 
\caption{\label{fig-mvoutput}
(color online) 
(a) NN$_H$ distribution for the $\mt0$ analysis at $m_H=165$ GeV;
(b) BDT distribution for the $\mtjj$ analysis at $m_H=150$ GeV; and
(c) BDT distribution for the $\etjj$ analysis at $m_H=150$ GeV.
The $t\overline t$, single top and diboson backgrounds are 
shown together as ``other".
}
\end{center}
\end{figure*}

The $\mt0$ analysis uses neural networks~\cite{tmva} (NN$_H$) trained 
to discriminate between all backgrounds
and all signals for $115 \leq m_H \leq 200$ GeV  in 5 GeV increments.  Type-2
$\tau$ samples are trained separately, while
the $\tau$ types 1 and 3 are combined for training to increase statistics.  The NN$_H$
distributions are binned in 21 equal sized bins for $0 < {\rm NN}_H < 1.05$.
The $\mtjj$ and $\etjj$ analyses use boosted decision trees (BDT)~\cite{tmva} trained for 
all signals against the sum of all backgrounds, with all
$\tau$ types combined
for Higgs boson masses $105\leq m_H \leq 200$ GeV in 5 GeV steps.  
The BDT output is binned in 15 bins spanning $-1 < {\rm BDT} < 1$ with a non-uniform
binning to assure sufficiently small statistical uncertainty in the predicted 
backgrounds within any bin.
We smooth the effects of signal MC statistics by averaging BDT distributions
for $m_H$ with the neighboring distributions at ($m_H-5$) GeV and ($m_H+5$) GeV
with weights of 50\%, 25\%, and 25\% respectively.
Figure~\ref{fig-mvoutput} shows the NN$_H$ distribution for 
the $\mt0$ analysis at $m_H=165$ GeV and the averaged BDT 
distributions for the $\ltjj$ analyses at $m_H= 150$ GeV, 
where the sensitivities are maximal.


\section{Systematic Uncertainties}

A large number of systematic uncertainties have been considered, typically
broken down separately for each analysis channel, tau type, background or signal process,
or Higgs boson mass.  The luminosity and trigger uncertainties are obtained from
separate analyses of D0 data.  The lepton, tau, and jet energy scale, resolution, and 
identification uncertainties are obtained from
special control samples.   Uncertainties
in the MC-simulated background cross section normalizations and shapes are obtained using theoretical
uncertainties, and the extent to which special data samples enriched in each background process
agree with MC predictions.  The MJ background uncertainties are determined
by comparing the alternate MJ-enriched samples with the results
obtained with the nominal choice.   Signal cross section uncertainties are obtained
from theoretical estimates and include the effect of PDF uncertainties.
For each source, the impact on the final variable (NN$_H$ or BDT) distribution
is assessed by changing the nominal values of a parameter by $\pm 1$ s.d.
Some of the uncertainties affect only the normalization of the final variable distribution
and some also modify its shape.

Table~\ref{tab-syst} summarizes the systematic uncertainties.  Many entries 
comprise several subcategories.  For example, the jet reconstruction uncertainty
includes the effects of jet identification, confirmation that the tracks within the
jet arise from the PV, jet resolution and jet energy scale.  Moreover,
these elements of the jet reconstruction uncertainty are computed separately for different background
processes and hypothesized Higgs mass values in each analysis channel.
The dominant systematic uncertainties are due to the $V+$jets and MJ backgrounds, with
significant contributions from jet reconstruction and modelling for the $\ltjj$ analyses.


\begin{table}[htbp]
\caption{\label{tab-syst}
The range of systematic uncertainties (in percent) for categories of their source.  
Each category generally
summarizes several individual sources separated by analysis channel, tau type, 
and/or physical process.  Those with ``Type" indicated as ``N'' affect only the
normalization of the final variable distribution.  Those indicated as ``S''
also affect the shape of the final variable distribution.
}
\begin{tabular}{lcc} \hline \hline
Source & Type & Uncertainty \\ \hline
Luminosity                       & N & $6.1$   \\
Muon trigger                     & N & $5-9$   \\
Electron trigger                 & N & $2$     \\
Muon reconstruction              & N & $2-3$   \\
Electron reconstruction          & N & $4$     \\
Tau reconstruction                & N & $4-14$  \\
Jet reconstruction               & S & $2-10$  \\
Jet modeling                    & S & $0-7$   \\
MC simulated backgrounds         & N & $5-12$  \\
MJ background                    & S & $10-50$ \\
Signal cross sections            & N & $5-40$  \\ \hline \hline
\end{tabular}
\end{table}

\section{Cross Section Limits}

We observe no excess of events over that expected from backgrounds in 
Fig.~\ref{fig-mvoutput}.  We therefore obtain
upper limits on the Higgs boson cross section for each analysis 
from the final multivariate outputs
using the modified frequentist method~\cite{modfreq}, using a 
negative log likelihood ratio (LLR) for
the background only and signal+background hypotheses as the test statistic.  For
the $\mt0$ analysis, each tau type is input separately to the limit setting 
calculation for Higgs boson masses from 115 to 200 GeV in 5 GeV steps.  The $\ltjj$ 
calculation uses the BDTs summed over tau type for $m_H$ values from 105 to 200 GeV in 5 GeV steps,
averaged over neighboring mass bins as described above.

The impact of systematic uncertainties on the limits is minimized by maximizing a 
likelihood function~\cite{wade} in which these uncertainties are constrained to Gaussian priors.
The value of the Higgs boson cross section is adjusted in each limit calculation until the value
of $CL_s$ reaches 0.05, corresponding to the 95\% C.L., where $CL_s = CL_{s+b}/CL_b$ and $CL_{s+b}$
($CL_b$) are the probabilities for the negative LLR value observed in 
simulated signal+background (background) pseudo-experiments to be less than that observed
in our data.  The limits obtained are summarized in Table~\ref{tab-limits}.


\begin{table}[t]
\caption{\label{tab-limits}
The ratio of expected and observed 95\% C.L. limits on the Higgs boson 
cross section to the SM values
for each analysis channel and the combination of all channels including that of~\cite{ttjj_old}.
}
\begin{tabular}{ccccccccc} \hline \hline
$m_H$ & \multicolumn{2}{c}{$\mt0$}  & \multicolumn{2}{c}{$\mtjj$} &
    \multicolumn{2}{c}{$\etjj$} &  \multicolumn{2}{c}{Combined} \\ \hline
 ~  & exp  & obs  & exp  & obs  & exp  & obs & exp  & obs   \\ \hline
105 & --   & --   & 17.7 & 18.5 & 33.3 & 58.9 & 12.6 & 17.1 \\
110 & --   & --   & 19.3 & 20.8 & 34.3 & 55.7 & 12.9 & 17.7 \\
115 & 84.2& 106.4 & 20.3 & 26.3 & 37.5 & 55.1 & 14.3 & 21.8 \\
120 & 42.9 & 31.1 & 19.2 & 23.3 & 40.5 & 59.4 & 13.7 & 15.6 \\
125 & 34.2 & 37.5 & 17.3 & 19.5 & 42.3 & 64.9 & 12.8 & 15.7 \\
130 & 25.2 & 32.4 & 15.9 & 20.6 & 44.2 & 72.5 & 11.5 & 17.9 \\
135 & 20.3 & 20.3 & 17.5 & 15.2 & 47.2 & 82.5 & 11.3 & 11.8 \\
140 & 16.7 & 20.0 & 18.7 & 13.2 & 44.7 & 68.1 & 11.1 & 10.1 \\
145 & 13.8 & 13.3 & 18.3 & 12.9 & 43.5 & 54.2 & 11.3 & ~9.8 \\
150 & 11.9 & 12.8 & 17.9 & 13.6 & 45.4 & 54.1 & 10.8 & ~9.5 \\
155 & ~9.8 & 12.9 & 18.2 & 13.2 & 42.3 & 57.5 & ~9.2 & ~9.0 \\
160 & ~8.2 & ~7.6 & 19.1 & 11.1 & 33.9 & 74.9 & ~8.4 & ~7.6 \\
165 & ~8.1 & ~7.8 & 21.7 & 11.2 & 32.8 & 69.8 & ~7.7 & ~6.8 \\
170 & ~8.5 & ~9.4 & 21.3 & 12.7 & 35.2 & 64.5 & ~8.5 & ~7.4 \\
175 & ~9.5 & ~8.6 & 22.7 & 11.4 & 40.7 & 73.7 & ~9.6 & ~8.0 \\
180 & 12.2 & 13.5 & 22.1 & 14.6 & 45.5 & 84.6 & 11.4 & 11.0 \\
185 & 13.5 & 12.1 & 25.7 & 19.8 & 53.7 & 90.8 & 12.2 & ~9.7 \\
190 & 16.5 & 17.2 & 29.5 & 19.1 & 58.8 & 101.8 & 14.6 & 12.3 \\
195 & 18.5 & 18.7 & 30.1 & 20.9 & 67.3 & 110.4 & 16.1 & 15.3 \\
200 & 19.2 & 31.5 & 28.9 & 26.9 & 69.3 & 114.4 & 19.8 & 29.9 \\ \hline \hline

\end{tabular}
\end{table}

\begin{figure}[htbp]
\includegraphics[scale=0.40]{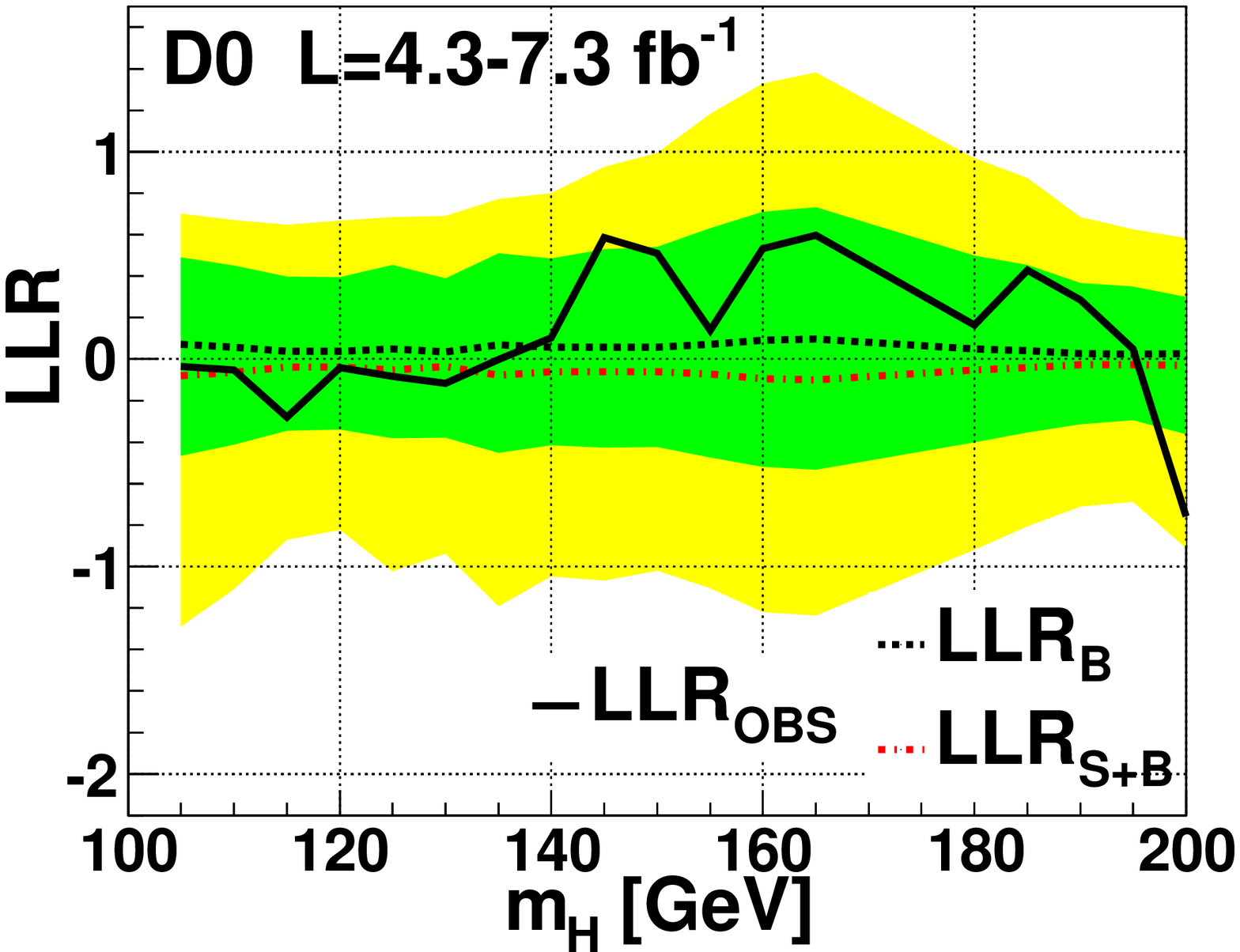}
\caption{\label{fig-llr}
(color online)
The LLR distribution as a function of $m_H$ showing 
the expected LLR distributions for the background only and signal+background hypotheses,
and the observed LLR, for the combination of all channels.
The green (yellow) bands show the $\pm 1$ s.d. ($\pm 2$ s.d.) bands around
the expected background only LLR values.
}
\end{figure}
\begin{figure}[htbp]
\includegraphics[scale=0.40]{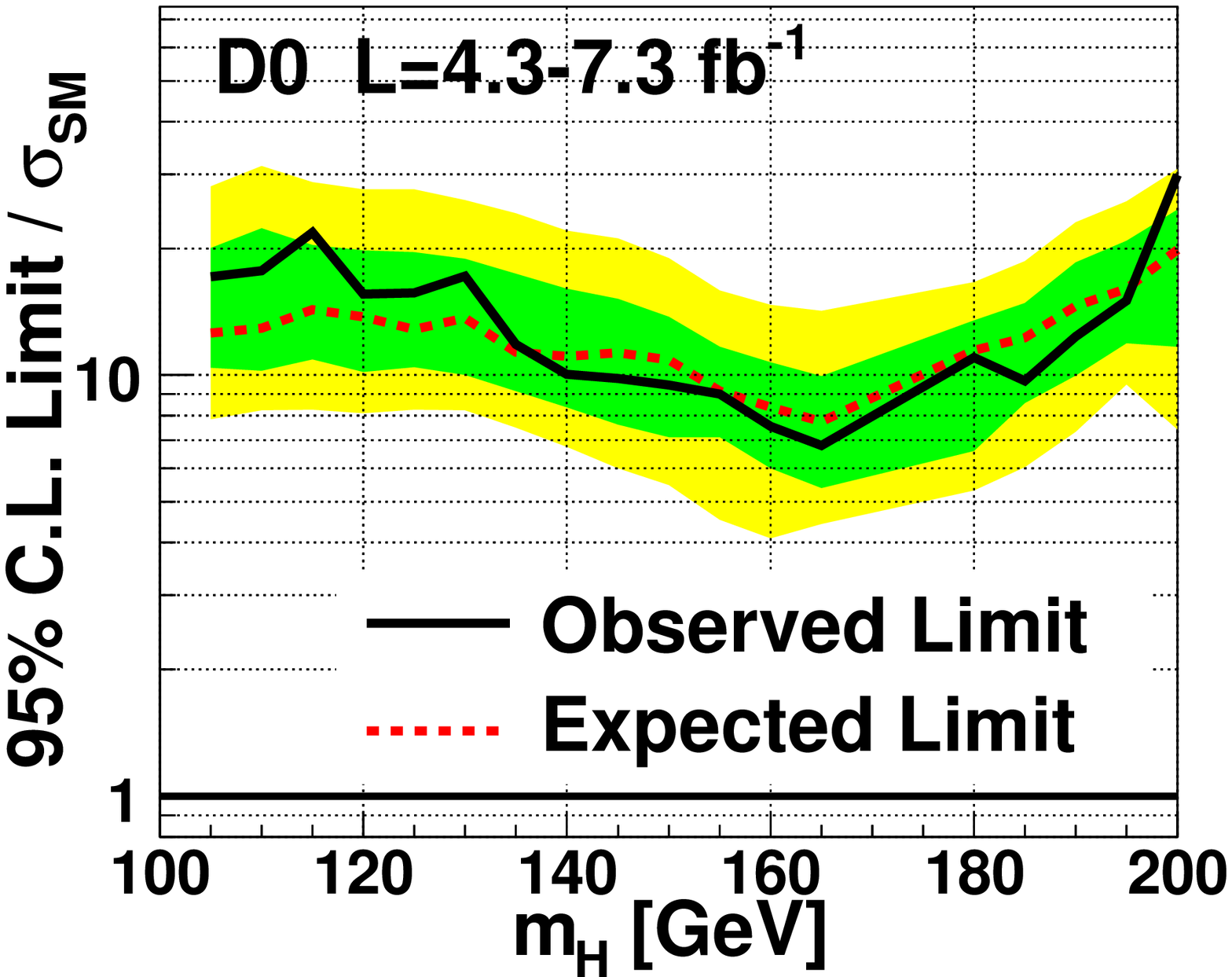}
\caption{\label{fig-limits}
(color online)
The ratio of observed and expected Higgs boson cross section limits to those expected in the SM,
for the combination of all channels. 
The green (yellow) bands show the $\pm 1$ s.d. ($\pm 2$ s.d.) bands around
the expected ratios.
  }
\end{figure}

We combine the information from the three channels by recomputing the LLR and limits
for the three analyses together, now also including the limits from the previous 
independent $\mtjj$ 
analysis using 1 fb$^{-1}$~\cite{ttjj_old}.  
In this calculation, the systematic uncertainties
across the different analyses are appropriately correlated (e.g. the $\zj$ normalization for
all channels is the same).  The fully combined LLR distributions and the 95\% C.L. limits as a function
of $m_H$ are shown in Figs.~\ref{fig-llr} and \ref{fig-limits}.  The combined limits are also shown in 
Table~\ref{tab-limits}.


In summary we have searched for the SM Higgs boson in final states involving an electron or muon
and a hadronically decaying tau.  We set 
95\% C.L. limits on 
the Higgs boson production cross section which are 21.8 and 6.8 times those expected in
the SM for Higgs boson masses of 115 and 165 GeV.

\vspace*{2.5mm}

%
We thank the staffs at Fermilab and collaborating institutions,
and acknowledge support from the
DOE and NSF (USA);
CEA and CNRS/IN2P3 (France);
MON, Rosatom and RFBR (Russia);
CNPq, FAPERJ, FAPESP and FUNDUNESP (Brazil);
DAE and DST (India);
Colciencias (Colombia);
CONACyT (Mexico);
NRF (Korea);
FOM (The Netherlands);
STFC and the Royal Society (United Kingdom);
MSMT and GACR (Czech Republic);
BMBF and DFG (Germany);
SFI (Ireland);
The Swedish Research Council (Sweden);
and
CAS and CNSF (China).


\vskip 16mm

\begin{thebibliography} {99}

\bibitem{lep-higgs}
R. Barate \etal~(LEP Working Group for Higgs boson searches), Phys. Lett. B {\bf 565}, 61 (2003).

\bibitem{tevatron-higgs}
T. Aaltonen \etal~(CDF and D0 Collaborations),
Phys. Rev. Lett. {\bf 104}, 061802 (2010).

\bibitem{lhc-higgs}
G. Aad \etal~(ATLAS Collaboration), Phys. Lett. B {\bf 710}, 49 (2012);
S. Chatrchyan \etal~(CMS Collaboration), Phys. Lett. B {\bf 710)}, 26 (2012).

\bibitem{gfitter}
M. Baak \etal~(GFitter collaboration), {\tt arXiv:hep-ph/1107.0975 (2011)},
(submitted to Eur. Phys. J. C).

\bibitem{newwmass}
With new measurements of the $W$ boson mass, T. Aaltonen \etal~(CDF Collaboration), 
Phys. Rev. Lett. 108, 151803 (2012)
and V.M.~Abazov \etal~(D0 Collaboration), 
Phys. Rev. Lett. 108, 151804 (2012), 
the constraints on $m_H$ from precision measurements are reduced by about 5\%.



\bibitem{ttjj_old} 
V.M.~Abazov \etal~(D0 Collaboration), Phys. Rev. Lett. {\bf 102}, 251801 (2009).

\bibitem{cdf_tautau}
T.~Aaltonen \etal~(CDF Collaboration), Phys. Rev. Lett. {\bf 108}, 181804 (2012). 

\bibitem{dzerodet}
S.~Abachi \etal~(D0 Collaboration), Nucl. Instrum. 
Methods Phys. Res. A {\bf 338}, 185 (1994);  
V.M.~Abazov \etal~(D0 Collaboration), Nucl. Instrum. 
Methods Phys. Res. A {\bf 565}, 463 (2006);  
V.M.~Abazov \etal , Nucl. Instrum. 
Methods Phys. Res. A {\bf 584}, 75 (2008);   
and V.M.~Abazov \etal , Nucl. Instrum. 
Methods Phys. Res. A {\bf 622}, 298 (2010).   


\bibitem{cteq}
J.~Pumplin \etal , J. High Energy Phys. {\bf 07}, 012 (2002).

\bibitem{alpgen}
M.L.~Mangano \etal , J. High Energy Phys. {\bf 07}, 001 (2003).

\bibitem{pythia}
T.~Sj\"ostrand, S.~Mrenna and P.~Skands, J. High Energy Phys. {\bf 05}, 026 (2006).

\bibitem{d0zpt}
V.M.~Abazov \etal ~(D0 Collaboration), Phys. Rev. Lett. {\bf 100}, 102002 (2008).

\bibitem{nnlowpt}
K. Melnikov and F. Petriello, Phys. Rev. D {\bf 74}, 114017 (2006).

\bibitem{nnlovjxs}
R. Hamburg, W.L. van Neerven, and W.B. Kilgore, Nucl. Phys. {\bf B359}, 343 (1991); 
{\bf B644}, 403 (2002).

\bibitem{mrst2004}
A.D.~Martin, R.G.~Roberts, W.J.~Stirling, and W.B.~Kilgore, Phys. Lett. B {\bf 604}, 61 (2004).

\bibitem{comphep}
E.~Boos \etal ,  Phys. Atom. Nucl. {\bf 69}, 1317 (2006); 
E.~Boos \etal~(CompHEP Collaboration), Nucl. Instrum. Methods Phys. Res. A
{\bf 534}, 250 (2004).

\bibitem{topxs}
N. Kidonakis, Phys. Rev. D {\bf74}, 114012 (2006); 
S.~Moch and P.~Uwer, Phys. Rev. D {\bf 78}, 34003 (2008).

\bibitem{higgsxs}
J. Baglio and A. Djouadi, 
J. High Energy Phys. {\bf 10}, 064 (2010).

\bibitem{hdecay}
A.~Djouadi, J.~Kalinowski, and M.~Spira, Comp. Phys. Commun. {\bf 108}, 56 
(1998).

\bibitem{tauola}
S.~Jadach \etal, Comp. Phys. Commun. {\bf 76}, 361 (1993).

\bibitem{geant}
R. Brun and F. Carminati, CERN Program Library Long Writeup W5013, 1993 (unpublished).

\bibitem{pv}
The primary vertex is that found to be the most likely collision point, among
possibly several collisions within a specific beam crossing, from which
our selected objects emanate.  The algorithm for defining it can be found in
V. M. Abazov et al. (D0 Collaboration), Nucl. Instrum.
Methods Phys. Res. A {\bf 620}, 490 (2010).

\bibitem{taunn}
V.M. Abazov \etal  (D0 Collaboration), Phys. Lett. B {\bf 670}, 292 (2009). 

\bibitem{jetalg} 
G. Blazey \etal ~in 
{\it Proceedings of the Workshop: QCD and Weak Boson Physics in Run II}, 
edited by U.~Baur, R.K.~Ellis and D.~Zeppenfeld, Fermilab-Pub-00/297 (2000).

\bibitem{nnel}
V.M.~Abazov \etal~(D0 Collaboration), Phys. Rev. Lett. {\bf 101}, 071804 (2008).

\bibitem{metsig}
A. Schwartzman, FERMILAB-THESIS-2004-21 (2004). 

\bibitem{madar}
R. Madar, FERMILAB-THESIS-2011-39 (2011). 

\bibitem{collinear-met}
D.~Rainwater, D.~Zeppenfeld, K.~Hagiwara, Phys. Rev. D {\bf 59}, 014037 (1999).

\bibitem{tmva} 
A. H\"ocker \etal, {\tt arXiv:physics/0703039v5} (2007).

\bibitem{modfreq}
A. Read, J. Phys. G: Nucl. Part. Phys. {\bf 28}, 2693 (2002);
T. Junk, Nucl. Instrum. Methods Phys. Res. A {\bf 434}, 435 (1999).

\bibitem{wade}
W. Fisher, FERMILAB-TM-2386-E (2007).

\end{thebibliography}
\end{document}